\documentclass{camera}
\usepackage{graphicx}
\usepackage{amsmath}
\usepackage{amssymb}
\usepackage{nicefrac}
\newcommand{\GGVren}{\langle \hspace{-0pt} \frac{\alpha_s}{\pi} G^2 \hspace{-0pt} \rangle}
\newcommand{\GGMren}{\langle \hspace{-0pt} \frac{\alpha_s}{\pi} \hspace{-2pt} \left( \frac{(vG)^2}{v^2} - \frac{G^2}{4} \right) \hspace{-0pt} \rangle}
\newcommand{\T}[1]{\text{T} \left[ #1 \right]}
\newcommand{\St}[1]{| #1 \rangle}
\newcommand{\sT}[1]{\langle #1 |}
\newcommand{\Gr}{\St{\Omega}}
\newcommand{\gR}{\sT{\Omega}}
\begin{document}
\title{Chiral symmetry and open-charm mesons\footnote{
Dedicated to the late Prof. Dr. I. Iori.}}
\author{T. Hilger \and B. K\"ampfer}
\organization{Forschungszentrum Dresden-Rossendorf, PF 510119,\\ 01314 Dresden, Germany\\
TU Dresden, Institut f\"ur Theoretische Physik, 01062 Dresden, Germany}

\maketitle

\begin{abstract}
Pseudo-scalar and scalar $D$ mesons are considered within
the QCD sum rule approach. We present an analysis of
the mass splitting of the pseudo-scalar $D - \bar D$ mesons
and the relation to QCD condensates.
Weinberg type sum rules are derived for  
chiral partners which highlight the role of the chiral condensate.
\end{abstract}

\vskip 3mm
\noindent {\bf 1. Introduction}
\vskip 3mm
Chiral symmetry and its breaking pattern represents an important
aspect of strong interaction physics governing to a large extent
the structure of hadrons. The spontaneous breaking of chiral symmetry
is signalled by the large value of the chiral condensate, 
$\langle \bar q q \rangle_0$, which is one quantity describing the
QCD vacuum. There is, furthermore, the breaking of the dilatation
invariance by quantum fluctuations leading to the non-zero value
of the gluon condensate. Further nonzero
vacuum expectation values of QCD operators characterize, among other
quantities, the QCD ground state and are of relevance for the
hadron spectrum. At nonzero density and temperature these condensates
may change and modify the hadronic excitations. For instance,
at baryon density $n$ and small temperature, the chiral condensate behaves as
$\langle \bar q q \rangle = \langle \bar q q \rangle_0 
(1 - \frac{\sigma_N n}{(m_u + m_d) \langle \bar{q} q \rangle_0})$ 
with $\sigma_N$ being the nucleon sigma term.
Accordingly, one
is looking for observables being sensitive to ''chiral restoration''
meaning effects indicating the drop of $\langle \bar q q \rangle$
with increasing density. 

Various experiments have been performed with the goal to seek for
modifications of hadron properties and to assign them to changes
of the QCD vacuum. Among 
currently running experiments are those by the HADES collaboration
\cite{HADES_exp}
investigating the dilepton emissivity of strongly interacting matter  
which is suspected to have some relation to chiral symmetry 
\cite{Rapp_Wambach}. It happens, however, that the relation is not
as direct as earlier conjectured \cite{Brown_Rho,Hatsuda_Lee,Ronny}.
One may look, therefore, for other potential observables with
more direct relations to the chiral condensate.
In \cite{Erice}, open charm mesons are envisaged as suitable
hadrons depending strongly on the chiral condensate.
Following this line of arguments we consider here QCD sum rules
for pseudo-scalar $D$ mesons (section 2) and address briefly 
sum rules for chiral partners (section 3). Such considerations
are related to the charm programmes at the forthcoming
experiments CBM \cite{CBM} and PANDA \cite{PANDA} at FAIR.
\vskip 3mm
\noindent {\bf 2. Spectral moments for $D - \bar D$ mesons}
\vskip 3mm
QCD sum rules \cite{QCDSR} represent a direct tool to relate moments
of hadronic spectral functions with QCD condensates via the 
operator product expansion (OPE). In \cite{Hilger}, the
following moments have been defined
\begin{equation}
S_n(M^2) \equiv \int_{s_0^-}^{s_0^+} ds \, s^n \, \Pi(s) \, {\rm e}^{-s^2/M^2} ,
\end{equation}
where $\Pi$ denotes the hadronic spectral function
and $s_0^\pm$ are the so-called continuum thresholds
bracketing the $D$ and $\bar D$ strengths.
Thereby, new quantities $\overline{\Delta m}$ and $\overline{m}$ 
may be defined which encode the combined mass-width properties 
of the particles under consideration:
\begin{equation}
\overline{\Delta m} \equiv  \frac12 \frac{S_1 S_2 - S_0 S_3}{S_1^2 - S_0 S_2} ,
\quad
\overline{m_+ m_-}  \equiv  - \frac{S_2^2 - S_1 S_3}{S_1^2 - S_0 S_2}
\end{equation}
and $\overline m^{\,2} \equiv \overline{\Delta m}^{\,2} + \overline{m_+ m_-}$.
For the often employed pole +continuum ansatz, these quantities
allow for an interpretation as mass splitting $2 \overline{\Delta m}$
and mass centroid $\overline{m}$.
 
Employing the current operators 
$\text{j}_{D^+} = i \overline{d} \gamma_5 c$ and $\text{j}_{D^-} = 
\text{j}^\dagger_{D^+} = i \overline{c} \gamma_5 d$, 
we obtain for the OPE evaluation of the current-current correlator
$ 
\Pi(q) = i \int d^4x \, e^{iqx} \gR \T{\text{j}(x) \text{j}^\dagger(0)} \Gr
$ 
up to mass dimension 5, 
in the rest frame of nuclear matter $v = (1, \vec{0} \,)$ 
($v$ stands for the medium four-velocity), in the limit of a light
$d$ quark mass $m_d \to 0$, 
for sufficiently large charm-quark pole mass $m_c$, and for the $D$ mesons at rest
the infrared-safe result \cite{Hilger} for the even ($e$) and odd ($o$)
Borel transformed correlators
\begin{align} \label{eq:borel_sr_d}
&\tilde \Pi^e( M^2)
=    \frac{1}{\pi} \int_{m_c^2}^\infty ds \, e^{-s/M^2} 
        \text{Im} \Pi_{D^+}^{per}(s, \vec{q} = 0)
        \\       
        &+ e^{-m_c^2/M^2} \left\{
        -m_c \langle \overline{d}d \rangle
        + \frac{1}{2} \left( \frac{m_c^3}{2M^4} - \frac{m_c}{M^2} \right)
        \langle \overline{d} g \sigma {\cal G} d \rangle
        + \frac{1}{12} \GGVren \right.
        \nonumber \\
        &+ \left[ \left( \frac{7}{18} + \frac{1}{3} \ln \frac{\mu^2 m_c^2}{M^4}
         - \frac{2}{3}\gamma_E \right)
        \left( \frac{m_c^2}{M^2} - 1 \right) - \frac{2}{3} \frac{m_c^2}{M^2}
        \right] \GGMren
        \nonumber \\      
        &+ \left.2 \left( \frac{m_c^2}{M^2} - 1 \right) \langle d^\dagger iD_0 d \rangle
        + 4 \left( \frac{m_c^3}{2M^4} - \frac{m_c}{M^2} \right)
        \left[ \langle \overline{d} D_0^2 d \rangle
        - \frac{1}{8} \langle \overline{d} g \sigma {\cal G} d \rangle \right]
        \right\} , \nonumber
    \\
&\tilde \Pi^o( M^2) =
e^{-m_c^2/M^2} \left\{
        \langle d^\dagger d \rangle
        - \left( \frac{2 m_c^2}{M^4} - \frac{4}{M^2} \right) \langle d^\dagger D_0^2 d \rangle
        - \frac{\langle d^\dagger g \sigma {\cal G} d \rangle} {M^2}
        \right\},
\label{eq:borel_sr_d_odd}
\end{align}
where $\alpha_s = g^2 /4 \pi$ and the perturbative spectral function 
$\text{Im} \Pi_{D^+}^{per}(s)$ is according to \cite{Aliev}.
We stress the occurrence of the term $m_c \langle \overline{d}d \rangle$,
where the large charm-quark mass acts as an amplifier 
of the genuine chiral condensate $\langle \overline{d}d \rangle$.

Results of the numerical analysis of (2) with
$S_0 = \tilde \Pi^o$, $S_1 = \tilde \Pi^e$ etc.\
are exhibited in Fig.~1 for condensates
listed in \cite{Hilger} with the exception of the poorly known
condensate $\langle q^\dagger g \sigma {\cal G} q \rangle$ for which we employ
here the opposite sign to get a measure of its importance. 
(This sign change 
also affects 
$\langle q^\dagger D_0^2 q \rangle$ due to
$\frac{1}{12} \langle q^\dagger g \sigma {\cal G} q \rangle - 
\langle q^\dagger D_0^2 q \rangle \approx 0.031 \text{ GeV}^2$.)
While the mass splitting is fairly robust, 
the centroid mass shift is fragile under variation 
of the continuum threshold parameters. The actual value of the
mass splitting indeed depends sensitively on the sign 
of $\langle q^\dagger g \sigma {\cal G} q \rangle$;
the other condensates 
in (\ref{eq:borel_sr_d}, \ref{eq:borel_sr_d_odd}) (cf.\ \cite{Hilger}) are fairly known.

\vskip 3mm
\noindent {\bf 3. Chiral partners with open charm} 
\vskip 3mm 
With the same technique one may derive difference QCD sum rules for
pseudo-scalar ($P$) and scalar ($S$) chiral partners, say
$D(J^P = 0^-)_{1864}$ and $D(J^P = 0^+)_{2400}$, with the result
\begin{equation} \label{eq:spmom}
\frac{1}{\pi} \int_{-\infty}^\infty ds \, s^i 
\left[ \Pi_{P-S}(s) \right] =  \langle {\cal O}_i \rangle
\end{equation}
with $\langle {\cal O}_i \rangle = -2 m_c \langle \overline{q}q \rangle$
for $i = 1$, $-2 m_c^3 \langle \overline{q}q \rangle 
+ m_c \langle\overline{q} g \sigma {\cal G} q \rangle$ for $i = 3$ and
$-m_c^5 \langle \overline{q}q \rangle 
+ \frac32 m_c^3 \langle\overline{q} g \sigma {\cal G} q \rangle + \cdots$
for $i = 5$, where $\cdots$ stand for condensates of mass dimension 7
and their Wilson coefficients.
These Weinberg type QCD sum rules clearly exhibit how the condensates
drive the difference between moments of spectral functions.

\begin{figure}[h]
\centerline{
\includegraphics[width=0.4\textwidth]{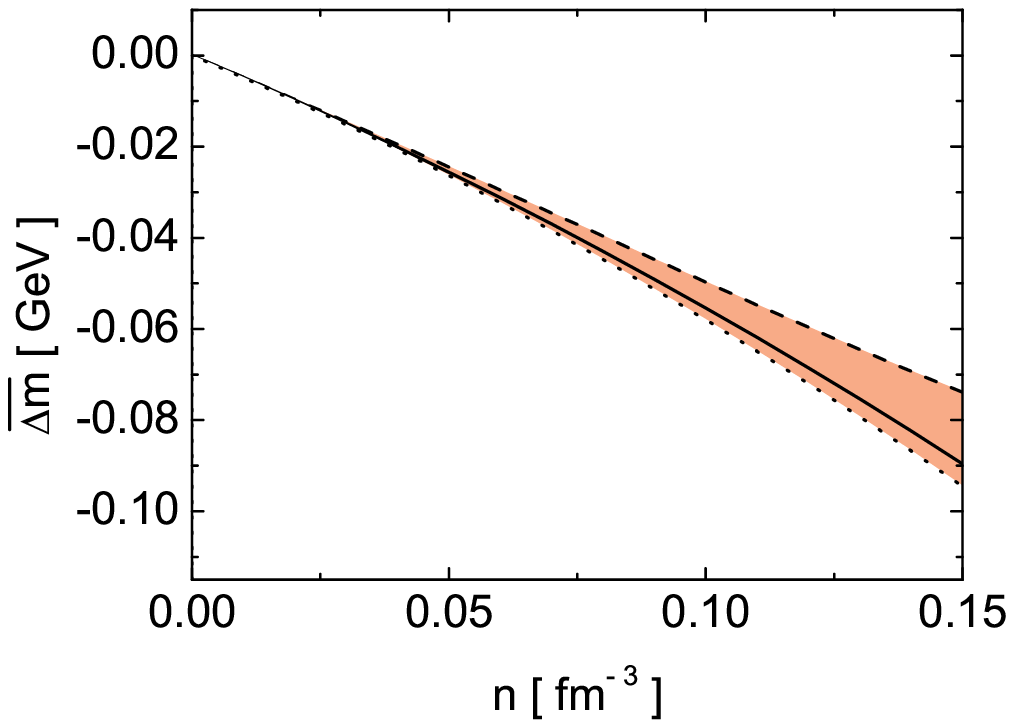}
\hspace*{1cm}
\includegraphics[width=0.4\textwidth]{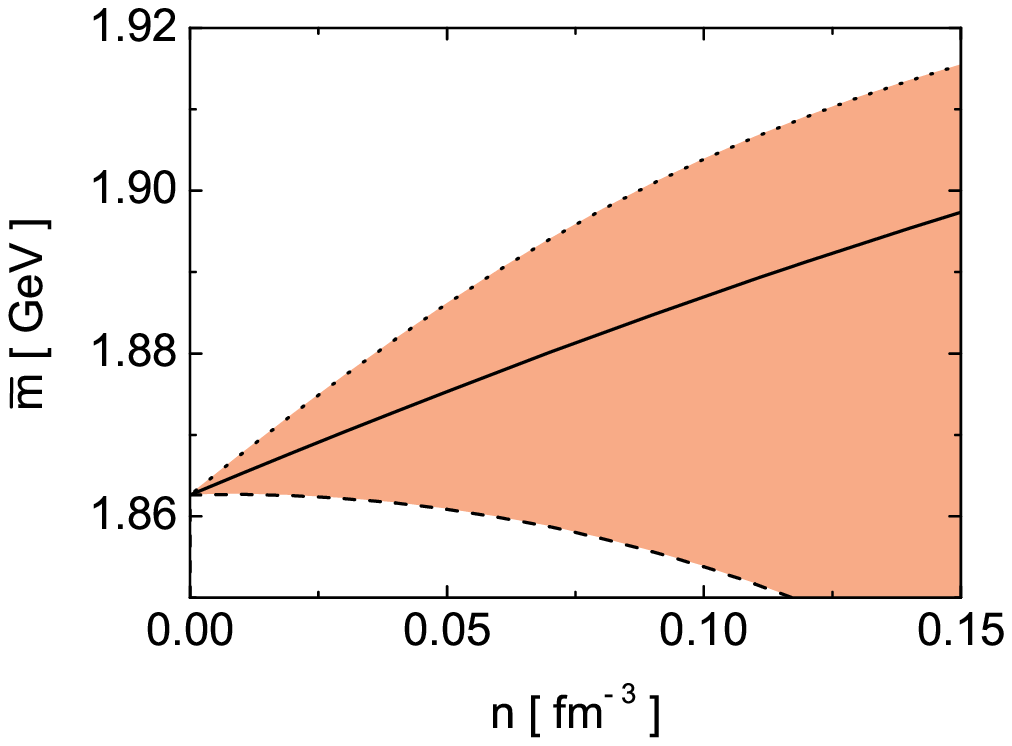}}
\caption{Evaluation of the QCD sum rules for the mass splitting (left)
and the centroid (right) of pseudo-scalar $D$ and $\bar D$ mesons for
$\langle q^\dagger g \sigma {\cal G} q \rangle = -0.33 \text{GeV}^2 \, n$.
The continuum thresholds are 
$s_0^{\pm\,2} = s_0^2 \pm \delta s_0^2$ with 
$s^2_0 = (6 + \xi n/n_0) \text{ GeV}^2$ 
(dotted, solid and dashed curves for $\xi = 1,\, 0$ and -1) and 
$\delta s_0^2$ according to \cite{Hilger}.}
\label{fig01}
\end{figure}

\vskip 3mm 
\noindent {\bf 4. Summary}
\vskip 3mm
In summary we have considered the QCD sum rule approach to pseudo-scalar
and scalar mesons composed of a light and heavy quark.
For the chiral sum rules the role of the chiral condensate is highlighted:
Dropping in-medium condensates, in particular $\langle \bar q q \rangle$, 
cause the degeneracy of the spectral moments of chiral partners.\\[1mm]
{\it Acknowledgements:} The work is supported by BMBF 06DR136 and GSI-FE. 

{\small
\begin{thebibliography}{99}
\bibitem{HADES_exp} HADES Collaboration (G. Agakishiev et al.), 
Phys. Rev. Lett. {\bf 98} (2007) 052302, Phys. Lett. B {\bf 663} (2008) 43, 
arXiv:0902.3478 [nucl-ex].
\bibitem{Rapp_Wambach} R. Rapp et al., Adv. Nucl. Phys. {\bf 25} (2000) 1;
arXiv:0901.3289 [hep-ph].
\bibitem{Brown_Rho} G.E. Brown, M. Rho, nucl-th/0509002, nucl-th/0509001. 
\bibitem{Hatsuda_Lee} T. Hatsuda, S.-H. Lee, Phys. Rev. C {\bf 46} (1992) 34. 
\bibitem{Ronny} R. Thomas, S. Zschocke, B. K\"ampfer, Phys. Rev. Lett. {\bf 95} (2005) 23230.
\bibitem{Erice} R. Thomas, T. Hilger, B. K\"ampfer, Prog. Part. Nucl. Phys. {\bf 61} (2008) 297.
\bibitem{CBM} http://www.gsi.de/fair/experiments/CBM/index\_e.html.
\bibitem{PANDA} http://www-panda.gsi.de/auto/phy/\_home.htm.
\bibitem{QCDSR} M.A. Shifman, A.I. Vainshtein, V.I. Zakharov, Nucl. Phys. B {\bf 147} (1979) 385. 
\bibitem{Hilger} T. Hilger, R. Thomas, B. K\"ampfer, Phys. Rev. C {\bf 79} (2009) 025202.
\bibitem{Aliev} T.M. Aliev, V.L. Eletsky, Sov. J. Nucl. Phys. {\bf 38} (1983) 936;\\
S. Narison, {\it QCD as a theory of hadrons}, 
Cambridge Monographs on Particle Physics, Nuclear Physics and Cosmology 17,
Cambridge 2004.
\end{thebibliography}}
\end{document}